\documentclass[12pt]{article}
\usepackage{epsf, cite, amssymb}
\usepackage{epsfig}
\usepackage{amsmath}
\makeatletter
\renewcommand\section{\@startsection {section}{1}{\z@}%
                                   {-3.5ex \@plus -1ex \@minus -.2ex}%nn
                                   {2.3ex \@plus.2ex}%
                                   {\normalfont\large\bfseries}}
\renewcommand\subsection{\@startsection{subsection}{2}{\z@}%
                                     {-3.25ex\@plus -1ex \@minus -.2ex}%
                                     {1.5ex \@plus .2ex}%
                                     {\normalfont\bfseries}}
\makeatother

\let\non\nonumber

\let\d=\delta

\let\s=\sigma

\let\x=\xi
\let\z=\zeta
\let\Th=\Theta

\newcommand{\bea}{\begin{eqnarray}}
\newcommand{\eea}{\end{eqnarray}}
\newcommand{\be}{\begin{equation}}
\newcommand{\ee}{\end{equation}}

\newcommand{\pref}[1]{(\ref{#1})}

\def\coeff#1#2{{\textstyle\frac{#1}{#2}}}
\newcommand{\hlf}{\frac{1}{2}}

\newcommand{\La}{\Lambda}
\newcommand{\la}{\lambda}
\newcommand{\T}{\theta}
\newcommand{\G}{\Gamma}

\newcommand{\e}{\epsilon}

\newcommand{\dd}{\delta}

\newcommand{\rr}{\rightarrow}
\newcommand{\m}{\mu}
\newcommand{\n}{\nu}
\newcommand{\p}{\partial}

\newcommand{\f}{\psi}

\newcommand{\xp}{x^+}

\newcommand{\lstr}{\ell_s}
\newcommand{\lpl}{\ell_p}

\newcommand{\C}[1]{$(\ref{#1})$}

\def\ie{{\it i.e.}}

\def\IZ{\relax\ifmmode\mathchoice
{\hbox{\cmss Z\kern-.4em Z}}{\hbox{\cmss Z\kern-.4em Z}}
{\lower.9pt\hbox{\cmsss Z\kern-.4em Z}} {\lower1.2pt\hbox{\cmsss
Z\kern-.4em Z}}\else{\cmss Z\kern-.4em Z}\fi}
\def\IR{\relax{\rm I\kern-.18em R}}

\def\one{{\hbox{ 1\kern-.8mm l}}}

\def\Tr{{\rm Tr\,}}

\newlength{\bredde}
\def\slash#1{\settowidth{\bredde}{$#1$}\ifmmode\,\raisebox{.15ex}{/}
\hspace*{-\bredde} #1\else$\,\raisebox{.15ex}{/}\hspace*{-\bredde}
#1$\fi}

\newsavebox{\zzzbar}
\sbox{\zzzbar}
  {\setlength{\unitlength}{0.9em}
  \begin{picture}(0.6,0.7)
  \thinlines
  \put(0,0){\line(1,0){0.6}}
  \put(0,0.75){\line(1,0){0.575}}
  \multiput(0,0)(0.0125,0.025){30}{\rule{0.3pt}{0.3pt}}
  \multiput(0.2,0)(0.0125,0.025){30}{\rule{0.3pt}{0.3pt}}
  \put(0,0.75){\line(0,-1){0.15}}
  \put(0.015,0.75){\line(0,-1){0.1}}
  \put(0.03,0.75){\line(0,-1){0.075}}
  \put(0.045,0.75){\line(0,-1){0.05}}
  \put(0.05,0.75){\line(0,-1){0.025}}
  \put(0.6,0){\line(0,1){0.15}}
  \put(0.585,0){\line(0,1){0.1}}
  \put(0.57,0){\line(0,1){0.075}}
  \put(0.555,0){\line(0,1){0.05}}
  \put(0.55,0){\line(0,1){0.025}}
  \end{picture}}

\newcommand{\ket}[1]{|{#1}\rangle}

\newcommand{\ena}{\end{eqnarray}}
\newcommand{\beqa}{\begin{eqnarray}}
\newcommand{\eeqa}{\end{eqnarray}}

\newcommand{\half}{\frac{1}{2}}

\def\G{\Gamma}

\newcommand{\g}{\gamma}

\def\d{\delta}
\def\e{\epsilon}
\def\g{\gamma}

\def\m{\mu}
\def\n{\nu}
\def\o{\omega}

\def\s{\sigma}

\def\x{\xi}
\def\z{\zeta}

\def\G{\Gamma}

\newcommand{\Dslash}{\ensuremath \raisebox{0.025cm}{\slash}\hspace{-0.32cm} D}

\begin{document}
\begin{titlepage}

\begin{center}

%{September 26, 2005}
March 14, 2006
\hfill         \phantom{xxx}

\hfill EFI-06-03

\vskip 2 cm
{\Large \bf Toward the End of Time}\\
\vskip 1.25 cm { Emil J. Martinec\footnote{email address: ejm@theory.uchicago.edu}, Daniel Robbins\footnote{email address:
robbins@theory.uchicago.edu} and Savdeep Sethi\footnote{email address:
 sethi@theory.uchicago.edu}}\\
{\vskip 0.5cm  Enrico Fermi Institute, University of Chicago,
Chicago, IL
60637, USA\\}

\end{center}
\vskip 2 cm

\begin{abstract}
\baselineskip=18pt

The null-brane space-time provides a simple model of a big crunch/big bang singularity. A non-perturbative definition of M-theory on this space-time was recently provided using matrix theory. We derive the fermion couplings for this matrix model and study the leading quantum effects. These effects include particle production and a time-dependent potential. Our results suggest that as the null-brane develops a big crunch singularity, the usual notion of space-time is replaced by an interacting gluon phase. 
This gluon phase appears to constitute the end of our conventional picture of space and time.

\end{abstract}

\end{titlepage}

\pagestyle{plain}
%\baselineskip=18pt
% Try a wider skip
\baselineskip=19pt
%%%%%%%%%%%%%%%%%%%%%%%%%%%%%%%%%%%%%%%%%%%%%%%%%%%%%%%%%%%%%%%%%%%%%%%%%%%%%%
\section{Introduction}

One of the major gaps in our understanding of string theory
is how it resolves space-like singularities.  A related issue
is how to formulate dynamics in time-dependent backgrounds
non-perturbatively.  The various
gauge/gravity correspondences subsume a resolution of
space-like singularities when they are localized inside 
black hole event horizons.  Implicitly, there must be
encoded the experience of a freely falling observer
who sees a time-dependent background culminating in
a space-like singularity.  One might then wonder whether
there is a regularization of cosmological singularities
via gauge theory.

Recently, some examples of such regularizations via gauge theory have been concretely 
described in~\cite{Craps:2005wd,Li:2005sz, Li:2005ti, Kawai:2005jx, Hikida:2005ec, Das:2005vd, Chen:2005mg, She:2005mt, She:2005qq, Chen:2005bk, Ishino:2005ru, Das:2006dr, Das:2006dz, Lin:2006ie}. These holographic models, based on matrix theory~\cite{Banks:1996vh}, appear to contain the right  degrees of freedom to describe physics at large blue shifts. In each case the semi-classical picture of the space-time singularity is replaced by a gluon phase. We will find further evidence for this picture in this work. Quantum effects in such models are quite fascinating have been examined recently in~\cite{Li:2005ai, Craps:2006xq}. 

Our goal is to develop this matrix approach when applied to M-theory in the particular background known as the null-brane space-time~\cite{Figueroa-O'Farrill:2001nx, Simon:2002ma}. The matrix model for this space-time was introduced recently in~\cite{Robbins:2005ua}. The null-brane is constructed using a particularly simple way of generating time dependence: namely, by orbifolding Minkowski space by 
a discrete subgroup of the Poincar\'e group.
This can yield a time-dependent background. For appropriate choices of orbifold group generators, this background can be singular. 

\begin{figure}[h] \label{figure1}
\begin{center}
\includegraphics[width=14cm]{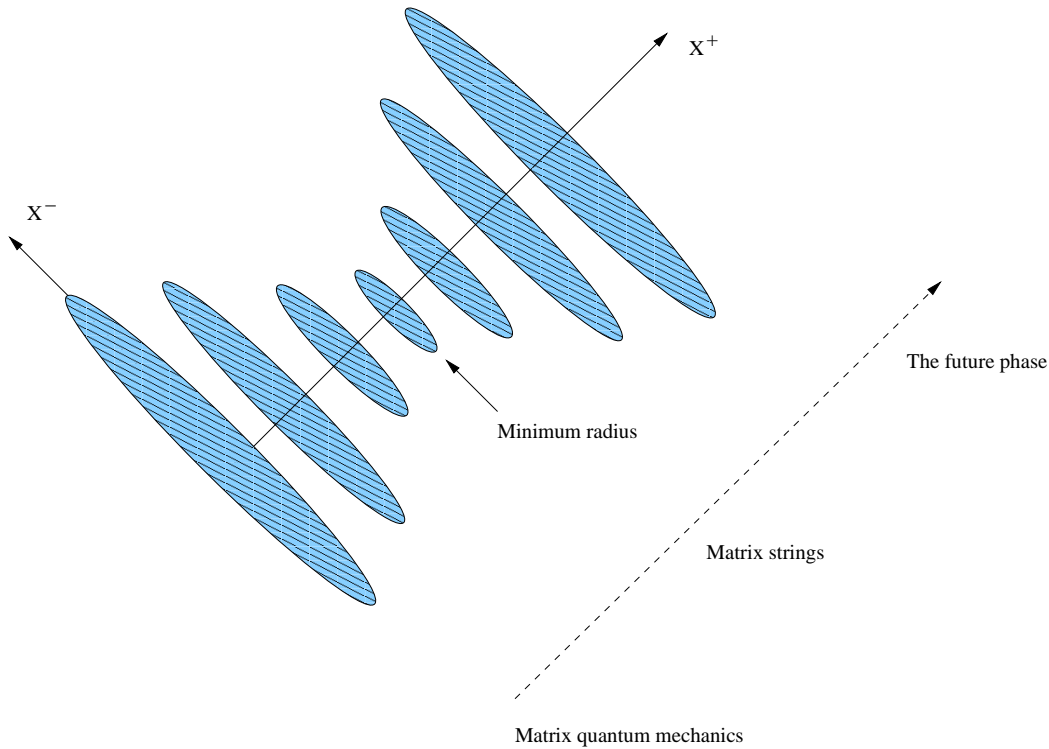}
\caption{\small The circle radius shrinks to a minimum size $L$ at $\xp =0$.}
\end{center}
\end{figure}

The null-brane is actually a non-singular solution of M-theory defined 
by choosing an $\mathbb{R}^{1,3}$ subspace of $\mathbb{R}^{1,10}$ with coordinates 
$$x^\pm = \frac{1}{\sqrt 2}(x^0 \pm
x^1), \, x, \, z, $$ 
and the usual flat metric $ds^2 = -2dx^+ dx^- +
dx^2 + dz^2$. We act on these coordinates by an element of the
4-dimensional Poincar\'e group
 \be \label{gen} g = \exp(2\pi i
K);\quad K = \frac{\lambda}{\sqrt 2}(J^{0x} + J^{1x}) + L P^z, \ee
where $L$ has dimensions of length. Under this action which depends on $(\lambda, L)$, \be
\label{quotgroup}
 X =
\left(\begin{array}{c}x^+\\x^-\\x\\z\\\end{array}\right)\
\rightarrow \quad
g\cdot X = \left(\begin{array}{c}x^+\\x^- + 2\pi \lambda x + 2\pi^2 \lambda^{2} x^+\\x + 2\pi \lambda x^+\\
z + 2\pi L\\\end{array}\right). \ee 
When $L$ goes to zero, the geometry has a null singularity at $x^+=0$. This space-time is depicted in figure~\ref{figure1}.

%The parameter $\lambda$ can be set to one by a light-cone boost
%\be\label{redef}
%x^{+} \rr {x^{+} \over \lambda}, \quad x^{-} \rr \lambda x^{-}. 
%\ee
%The space-time theory therefore depends only on the Planck length $\ell_{p}$ and the shift length $L$. 

{}For $L$ large compared to the Planck scale, 
this is simply a smooth time-dependent background 
of M-theory or string theory with a well-defined S-matrix. 
String perturbation theory on this background has been studied 
in~\cite{us, Liu:2002kb, Fabinger:2002kr} while the semi-classical stability 
of this space-time has been examined in~\cite{Lawrence:2002aj, Horowitz:2002mw}. 
Not surprisingly, string perturbation theory breaks down as $L\rr 0$.  
Holographic descriptions of string cosmologies, in the spirit of AdS/CFT, 
have been studied in~\cite{Elitzur:2002rt, holo, holo2}. 
In the case of the null-brane, these descriptions involve space-time non-commutative Yang-Mills theories.    

By contrast, the matrix description is obtained by performing 
an additional light-like compactification $x^- \sim x^- + 2\pi R$ which commutes with
the orbifold identification~\C{quotgroup}. This
results in a discrete light-cone quantized description of the theory.   
In close analogy to the flat space case~\cite{Seiberg:1997ad}, there is a decoupling limit~\cite{Robbins:2005ua}\ that reduces the matrix model to the low-energy dynamics of D0-branes 
moving in the orbifold~\pref{quotgroup}. 

The matrix theory description can also be thought of
as viewing the system from the perspective of
a highly boosted frame.  It does not directly 
describe the space-time vacuum, but rather describes 
objects in space-time that carry large $p^+$,
and therefore preserve at most one-half of the supersymmetries.
In the case of the null-brane, the one-half of the supersymmetries
preserved by the orbifold identification are incompatible with
the one-half of the supersymmetries preserved by boosting along
the light-cone direction.  Thus the null-brane matrix model
preserves no supersymmetry. We will see this explicitly when we evaluate the one-loop effective potential of the model, and compute the amplitude for particle production.    

We can understand the dynamics of a D0-brane in this orbifold background from the perspective of the covering space of the orbifold group action~\C{quotgroup}. On the covering space, we see a D0-brane together with an array of boosted images constrained to move in a manner invariant under the identification~\C{quotgroup}. The interactions between the image branes arise from the usual velocity-dependent forces between D0-branes. 

Semi-classically, when the brane and enough of its images are contained within their Schwarzschild radius, a black hole forms. For finite $L$, the black hole eventually decays and we can compute a conventional S-matrix.  However, as $L$ becomes small, this black hole becomes larger eventually filling all of space~\cite{Horowitz:2002mw}. We will find a quite different picture of this process from the matrix model which, in particular, includes degrees of freedom not seen in the semi-classical picture.    

T-dualization of this
system of D0-branes along the $z$ direction circle
yields a D1-brane matrix string theory description. The evolution from matrix quantum mechanics to matrix strings near the singularity is also depicted in figure~\ref{figure1}. Physics near the singularity is controlled by $1+1$-dimensional field theory dynamics rather than quantum mechanics. In principle, the future phase is a return to matrix quantum mechanics as the z circle becomes large again (and the T-dual field theory circle becomes small). 

However, strong quantum effects near the singularity can drastically alter the future description. Determining these quantum effects at one-loop in the matrix model is our primary task. We will find that a potential is generated in the matrix model which turns off at early and late times. This supports the consistency of the model since we need exact flat directions in the far past and future to recover a semi-classical picture of space-time.  

\begin{figure}[h] 
\begin{center}
\[
\mbox{\begin{picture}(100,170)(100,120)
\includegraphics[scale=.5 ]{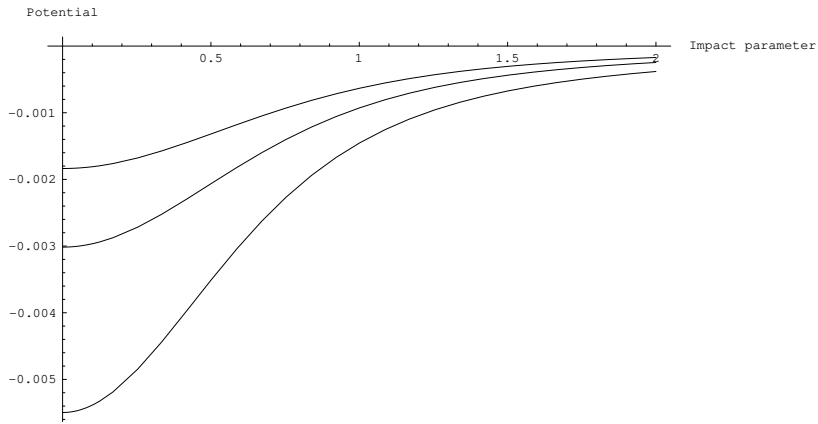}
\end{picture}}
\]
\caption{{\small The potential given in~\C{finalpot}\ as a function of impact parameter with various choices of $L$. The smaller $L$, the deeper the potential well. The choice of parameters is described in section~\ref{nonabelian}. }}\label{mygr2}
\end{center}
\end{figure}

The potential tends to attract gravitons more and more strongly as they approach $x^{+}=0$ and as we tune $L \rr 0$. This effective potential as a function of the impact parameter for two gravitons is displayed in figure~\ref{mygr2}. As the gravitons approach each other, the off-diagonal matrix degrees of freedom become lighter and our notion of space-time becomes fuzzy and is eventually replaced by matrix dynamics. At this level of approximation, it appears that there is no escape from this gluon phase as $L\rr 0$.   

The attraction of gravitons as $x^+$ and $L$
go to zero is a reflection of their interaction
 with one another's images under the orbifold group.
In fact, even a single graviton interacts with its own images
and can generate a spacelike singularity
at the null-brane neck at $x^{+}=0$;
even if the vacuum does not evolve to a singularity,
excited states can do so.  The null-brane matrix model 
describes objects (such as gravitons)
carrying longitudinal momentum in the null-brane background,
rather than being a description of the null-brane vacuum directly.
Thus even the ground state of the matrix model can evolve to a singularity
when $L$ is small enough.  This property is related to the lack of
supersymmetry of the null-brane matrix model.

The formation of such singularities by excitations on
the null-brane was analyzed by Horowitz and Polchinski from
the point of view of the dual gravitational theory in~\cite{Horowitz:2002mw}.
They argued that a spacelike singularity will form in the null-brane orbifold
whenever the gravitational radius of excitations exceeds
the proper size $L$ of the null-brane neck at $x^{+}=0$.
For $d>4$ noncompact spatial dimensions, the singularity 
that results is a finite-sized black hole;%
\footnote{More precisely, a black string which fills the $z$ direction.
After the neck is passed and the size of the $z$ circle starts to grow,
a Gregory-Laflamme instability sets in~\cite{Gregory:1993vy}
and this black string breaks up into one or more black holes.}
outside the black hole there is no big crunch singularity.
The gravitational radius of the black hole is
\be
\label{gravrad}
r_0=\Bigl(\frac{G_{\! N}\, p^+}{L^2}\Bigr)^{\frac1{d-4}}
\ee
in terms of the effective $d+1$ dimensional Newton constant $G_N$
and momentum $p^+$ of the initial excitation.  In particular, 
the size of the black hole grows to infinity as $L\to 0$.
For $d=4$, no black hole forms for large $L$, and an infinite mass
black hole forms for small $L$.  For $d<4$, a spacelike singularity
of infinite extent always forms.

The interesting regime of $d<4$ 
is outside the reach of matrix theory.
Matrix theory constructions describe gravity in terms of a dual field theory
for $d> 5$, and a dual ``little string theory'' for $d=5$; there are no
constructions for $d< 5$ (because the growth of the density
of states is too rapid to be described by 
a field theory or nongravitational string theory).
Thus, for finite $L$, matrix theory is only capable of describing the regime where 
the outcome of objects falling through the neck of the null-brane
is a finite-sized black hole.  The size of that black hole grows to infinity 
as $L/\lpl$ goes to zero.  One might view the matrix theory as a description
of the cosmological horizon that forms in this limit.

Black holes are well-approximated in matrix theory at finite $N$
provided their entropy is less than $N$~\cite{Banks:1997hz,Horowitz:1997fr,Li:1998ci}.  
Therefore, the
growing size of the final state black hole with decreasing $L$
puts a lower bound on $N$
\be
N_{\rm min} \sim \Bigl(\frac{\lpl^3 p^+}{L^2}\Bigr)^{\frac{d-1}{d-4}}
\ee
in order to have the matrix model approximate well the bulk
gravitational physics.  A conservative order of limits 
sends $N\to\infty$ first, then $L\to 0$ to achieve a cosmological
singularity.  The dynamics may sensitively depend on this order
of limits, however an intriguing possibility is that one could obtain
a useful picture of the cosmological singularity by taking $L\to 0$
for finite $N$ and then subsequently the large $N$ limit.
The $L\to 0$ limit of the finite $N$ theory accesses the 
perturbative regime of the gauge theory; if this order 
of limits is germane,
the cosmological singularity devolves to a kind
of free gluon phase.  If on the other hand the order of limits
does not commute, one will need to contend with the strong
coupling, large $N$ physics of the gauge theory;
still, the matrix model seems to provide a resolution
of the cosmological singularity.

This picture of the null brane
 is very much in accord with the model studied in~\cite{Craps:2005wd, Craps:2006xq}\ 
 in which time begins in a free gluon phase.  While our understanding 
 of the gluon phase (free or otherwise) in the null-brane is still quite preliminary 
 with much to be understood, it is exciting that the matrix model 
 appears to be a complete description containing all the ingredients necessary 
 for understanding the fate of the S-matrix and, indeed, the fate of space and time.

%%%%%%%%%%%%%%%%%%%%%%%%%%%%%%%%%%%%%%%%%%
%%%%%%%%%%%%%%%%%%%%%%%%%%%%%%%%%%%%%%%%%%

\section{The Matrix Action}

Our first task is to determine the complete action for the matrix description of M-theory on the null-brane. The bosonic couplings were determined in~\cite{Robbins:2005ua}, but we also require the fermion couplings to compute quantum corrections. 

\subsection{Parameters, scales and sectors}
\label{parameters}

It is important to keep track of the scales appearing in this system. 
The space-time physics depends on the matrix theory parameters 
$(\ell_{p}, R)$ as well as the orbifold parameters $(L,\lambda)$. 
The former are related to string theory parameters $(\ell_s, g_s)$ via
\be
\lpl = g_s^{1/3}\lstr, \quad\qquad R = g_s\lstr. 
\ee
We will express the matrix action in terms of string parameters and set $\ell_{s}=1$ for simplicity.  
The bosonic couplings in the matrix model describing M-theory on the null-brane with $p^{+}=N/R$ are given by a $1+1$-dimensional $U(N)$ gauge theory with action~\cite{Robbins:2005ua}
\bea\label{conjecture}
S &=& \frac{1}{ g_s}\int d\tau d\s\, \Tr\left(\vphantom{\frac{i\tau}{\sqrt 2}}(D_0x^i)^2 +(D_0x)^2-L^2 (D_1x^i)^2 -L^2 (D_1x)^2  - \sqrt{2} \lambda (x - \tau D_0x) F \right.\non\\
&&\hskip 3cm
 \left.+ (L^2+\hlf \lambda^{2}\tau^2) F^2 + {1\over 2} \left[ x^i, x^j\right]^2 + ( \left[x,x^i\right] +{i\tau \lambda \over \sqrt{2}} D_1x^i)^2 \right), 
\eea
where $ 0 \leq \sigma \leq 2\pi$. 
Here $\tau$ can be identified with the coordinate $x^+$ and $\sigma$ 
is a coordinate T-dual to $z$. 

There are $7$ scalar fields $x^{i}$ rotated by an $SO(7)$ symmetry. There is an additional scalar  field $x$ distinguished by the orbifold identification~\C{quotgroup}\ and a $U(N)$ gauge field with field strength $F$. Each of these adjoint-valued fields has canonical mass dimension $1$ while $\sigma, \tau$ and $L$ have length dimension $1$.  These are the natural dimension assignments from the perspective of gauge theory. Lastly, we note that as $\lambda \rr 0$, we recover standard matrix string theory~\cite{Motl:1997th, Banks:1996my, Dijkgraaf:1997vv}. 

The singularity that develops in space-time as $L \rr 0$ is reflected classically in the matrix model by the appearance of a new flat directions at $\tau=0$. These flat directions correspond to $\sigma$ fluctuations of the $x^{i}$ fields which are suppressed by the couplings, 
\be
(L^2+\hlf \lambda^{2}\tau^2 ) (D_{1} x^{i})^{2}, 
\ee
which vanish at $L=\tau=0$. We will examine how this classical picture is modified quantum mechanically. 

This  model can also be extended to describe string theory on the null-brane, or more generally M-theory on products of tori with the null-brane. This extension is straightforward and involves promoting some of the $x^{i}$ to gauge-fields while increasing the dimension of the field theory in the usual way~\cite{Taylor:1996ik}
\be
x^{i} \rr A^{i}. 
\ee
These extensions are already interesting for the case of string theory on the null-brane. Type IIA on the null-brane involves an extension of type IIB matrix string theory~\cite{Sethi:1997sw, Banks:1996my}\ in which time-dependent non-perturbative field theory effects should play an important role. On the other hand, type IIB on the null-brane involves a $3+1$-dimensional time-dependent gauge theory generalizing~\cite{Susskind:1996uh, Ganor:1996zk}. The action of S-duality in this theory should be quite fascinating. We will not pursue these directions here. 

Lastly we note that the M-theory on the null-brane has superselection sectors specified by the quantized amount of graviton momentum on the z circle. This circle is large at infinity so these excitations are finite energy. Each sector with non-zero momentum can be studied in the matrix model by analyzing the quantum dynamics expanded around a classical background with non-zero electric flux.       

\subsection{Decoupling revisited and the large $N$ scaling}
\label{largeN}
The decoupling limit for the null-brane matrix model can be understood along the lines of the original argument given in~\cite{Banks:1996vh}. Let us return to the orbifold perspective where we consider an array of D0-branes to make the connection between the arguments and determine how to take the large $N$ limit. 

The D0-brane action in string units has the schematic form
\be
\label{dzeroact}
  {\cal S} = \frac{1}{g_s\lstr}\int \!dt\,\Bigl( (\partial_t x)^2+[x,x]^2/\lstr^4
  +{\rm fermions}\Bigr)\ .
\ee
The matrix theory limit sends the dimensionless parameter $g_s$ to zero,
holding fixed the dimensionless quantities
\be
\hat t=g_s^{1/3}t/\lstr=Rt/\lpl^2, 
\quad\qquad 
\hat x = g_s^{-1/3}x/\lstr = x/\lpl. 
\label{scalinglimit}
\ee
The action expressed in terms of these parameters refers only to quantities in the resulting 
eleven-dimensional theory
\be
{\cal S} = \int\! d\hat t\, \Bigl[\Bigl(\frac{\partial\hat x}{\partial\hat t}\Bigr)^2+[\hat x,\hat x]
+{\rm fermions}\Bigr]\ .
\ee
At these scales the interactions are of unit strength.

We now come to the issue of scaling the parameters $(L,\lambda)$
of the null-brane orbifold.  First of all, we should hold fixed $L/\lpl$,
the size of the neck in the geometry at $\tau=0$ in eleven-dimensional
Planck units.  The scaling of $\lambda$ is can be understood
from the underlying D0-brane picture which is related to the 
1+1 QFT appearing in~\pref{conjecture}\ as a T-dual representation.

In the D0-brane picture, one has groups of D0-branes 
in relative motion on different fundamental domains of the covering
space of the orbifold.
Strings stretching between image branes are represented by
the Fourier modes of the 1+1 field theory.
The D0-brane representation is, however, useful for understanding
the matrix theory scaling limit of
the null brane orbifold.
The kinetic energy of relative D0-brane motion is excess energy above
the BPS threshold 
\be
 E_{\rm YM} = E-N/R = \sqrt{(N/R)^2+\vec p^2}-N/R = \frac{\vec p^2}{2N/R}+\dots
\label{YMenergy}
\ee
where the subleading terms in the last expression
vanish in the scaling limit.  The time scale \pref{scalinglimit}
means the energy in the D0-brane dynamics should scale in order to keep
$E_{\rm YM}\lstr g_s^{-1/3}$ fixed.  Equivalently, we are holding fixed
\be
\lpl^2\vec p^2 = E_{\rm YM} \cdot 2 N(\lpl^2 /R) \ .
\label{invtmass}
\ee
Now, D0-branes on neighboring images of the covering space
have relative kinetic energy
\be
\lpl^2(p-g\cdot p)^2 = (2\pi \lambda\lpl/R)^2\ .
\label{relKE}
\ee
In order to hold this quantity fixed, 
$\lambda$ must scale as $\lambda\sim g_s^{2/3}$.
This is the scaling limit proposed in~\cite{Robbins:2005ua}.  
In particular, standard matrix theory
arguments for the decoupling of both higher string modes and ten-dimensional
gravitational physics may be directly carried over 
to the null-brane matrix model.

More precisely, decoupling holds 
as long as one takes the scaling limit first holding fixed 
\be 
L/\lpl,  \quad \lpl^2 E_{\rm YM}/R, \quad  \lambda/g_s^{2/3}, \ee 
for fixed $N$.
One can then contemplate further taking $L/\lpl$ to zero, or $N$ to infinity,
in various combinations.  

Consider now the large $N$ limit.
In order to hold the invariant mass~\pref{invtmass} fixed
as we scale $N$, the Yang-Mills energy $E_{\rm YM}$
should scale as $1/N$.  Since there are $N$ D0-branes,
the kinetic energy $v^2/R$ of each must scale as $1/N^2$,
or $v\sim 1/N$ (\ie\ the velocity slows canonically
as one boosts the frame by dialing $N$).
The relative velocity of D0-branes related by $g^n$ 
is $2\pi n\lambda$; thus $\lambda\sim 1/N$.

\subsection{Fermion couplings from the quotient action}

To derive the fermion couplings, we can use the orbifold description of the null-brane. We will begin by studying the Euclidean instanton problem and then extend the result to D0-branes moving on the null-brane. The Euclidean instanton problem was also studied in~\cite{Berkooz:2005ym}.    

The supersymmetric instanton action in flat space is given by
\be\label{instaction}
S=- \Tr\left(\frac{1}{4}\left[X_\m,X_\n\right]\left[X^\m,X^\n\right]- \bar\Th \g_\m\left[X^\m,\Th \right]\right),
\ee
where $X^\m$  are bosons with $\m=0,\ldots,9$. The fermion, $\Th$, is the dimensional reduction of a Majorana-Weyl spinor in ten dimensions. Both $X^{\m}$ and $\Th$ are in the adjoint representation of the gauge group. We will take the gauge group to be $U(N)$. For the moment, we will take all fields to be dimensionless.  

We can express the action of the null-brane quotient group on the $X^{\m}$ in the following way
\be
X \, \rr \,\exp(\omega) X + a
\ee
where
\be
\omega^\m\vphantom{\omega}_\n=\eta^{\m\rho}\omega_{\rho\n},
\ee
and
\be
\omega_{x+}=-\omega_{+x}=2\pi\la,\qquad a^z=2\pi L,
\ee
with all other components vanishing.

The actions on the fermions can then be easily computed.  The translations have no effect on a spinor representation, while the Lorentz transformation acts via
\be
\Th \, \rr \, \exp\left(\frac{1}{4}\omega_{\m\n}\g^{\m\n}\right)\Th =\left(1-\pi\la\g^{+x}\right)\Th.
\ee
Note that half of the fermions, those that are annihilated by $\g^+$, are invariant under the quotient action.  This is equivalent to the statement that the background preserves one-half of the supersymmetries.

%%%%%%%%%%%%%%%%%%%%%%%%%%%%%%%%%%%%%%%%%%

\subsubsection{Instantons on the quotient space}

We are now ready to study instantons on the quotient. This should really be viewed as a formal exercise on route to determining couplings for dynamical D-branes on the null-brane.   So we view $X^{\m}$ and $\f$ as infinite matrices satisfying the constraints imposed by invariance under the quotient action
\bea
X^{+,i}_{m+k,n+k} &=& X^{+,i}_{mn},\non\\
X_{m+k,n+k} &=& X_{mn}+2\pi k\la X^+_{mn},\non\\
X^-_{m+k,n+k} &=& X^-_{mn}+2\pi k\la X_{mn}+2\pi^2k^2\la^2X^+_{mn},\\
Z_{m+k,n+k} &=& Z_{mn}+2\pi kL\dd_{mn},\non\\
\Th_{m+k,n+k} &=& \left(1-\pi k\la\g^{+x}\right)\Th_{mn}.\non
\eea
As usual, we Fourier transform this system (see~\cite{Robbins:2005ua}\ for this procedure applied to the bosons in this case). The matrix $\Th_{mn}$ is replaced by the operator
\be\label{Fourier}
\Th(\s,\s')=\sum_{m,n}e^{ i(n\s'-m\s)}\Th_{mn}=\left[\T(\s)-\frac{i\la}{2}\g^{+x}\T(\s)\p_\s\right]\,\dd(\s-\s')
\ee
where $ 0 \leq \s  \leq 2\pi$. The function $\T(\s)$ appearing in~\C{Fourier}\ is not Hermitian but this can be corrected by defining
\be
\tilde\T=\T+\frac{i\la}{4}\g^{+x}\T',
\ee
where the prime denotes a $\s$-derivative. Expressed in terms of $\tilde\T$,  the operator $\Th$ is Hermitian:
\be
\Th=\tilde\T-\frac{i\la}{4}\g^{+x}\tilde\T'-\frac{i\la}{2}\g^{+x}\tilde\T\p_\s.
\ee
Under a gauge transformation, $\tilde\T$ transforms according to the rule
\be
\tilde\T\rr u\tilde\T u^\dag+\frac{i\la}{4}\g^{+x}\left(u'\tilde\T u^\dag-u\tilde\T u^{\dag\prime}\right)
\ee
where $u(\s)$ is an element of $U(N)$ for the case of $N$ instantons. This is an unconventional gauge transformation rule. With $L\ne 0$, we can define a natural gauge covariant fermion, 
\be
\hat\T=\tilde\T+\frac{\la}{4L}\g^{+x}\left\{z,\tilde\T\right\}.
\ee
If we also define the covariant derivative, $D_1=\p_\s-iL^{-1}z$, then we see that
\be
\Th=\hat\T-\frac{i\la}{4}\g^{+x}\left\{D_1,\hat\T\right\}
\ee
which is manifestly Hermitian and covariant. 

We can combine this expression with the analogous operator expressions for the bosons found in~\cite{Robbins:2005ua}\ 
\bea\label{bigoperators}
& & X^{+,i} = x^{+,i}, \non\\
&&X = \hat x+\frac{i\la}{2}\left\{x^+,D_1\right\}, \non\\
&& X^- = {\hat x}^-+\frac{i\la}{2}\left\{\hat x,D_1\right\}-\frac{\la^2}{2}D_1\cdot x^+\cdot D_1, \non\\
&& Z = iLD_1,
\eea
to compute the Yukawa coupling appearing in~\C{instaction}
\bea\label{yukawa}
\Tr\left(\bar\Th\g_\m\left[X^\m,\Th\right]\right) &=& {1\over 2 \pi} \int d\s\, \Tr\left(\vphantom{\frac{\la^2}{8}}iL \bar\T \g^zD_1\T+\bar\T \left[-\g^-x^+-\g^+{\hat x}^-+\g^x\hat x+\g^ix^i,\T\right]\right.\non\\
&& \left.+\frac{i\la}{4}\left(\bar\T \g^{+-x}\left\{\T,D_1x^+\right\}+\bar\T \g^{+xi}\left\{\T,D_1x^i\right\}-D_1\bar\T\g^x\left\{\T,x^+\right\}\right.\right.\non\\
&& \left.\left.+\bar\T \g^x\left\{D_1\T,x^+\right\}+D_1\bar\T \g^+\left\{\T,\hat x\right\}-\bar\T \g^+\left\{D_1\T,\hat x\right\}\right)\right.\non\\
&& \left.+\frac{\la^2}{8}\left(D_1\bar\T \g^+\left[D_1\T,x^+\right]-D_1\bar\T \g^+\left[\T,D_1x^+\right]\right.\right.\non\\
&& \left.\left.-\bar\T \g^+\left[D_1\T,D_1x^+\right]\right)\vphantom{\frac{\la^2}{8}}\right).
\eea
We have dropped the hats on each $\T$ appearing in~\C{yukawa}\ to avoid notational clutter. Some of the terms above are not manifestly Hermitian but this is not problematic since they can be made Hermitian by adding total derivatives. We will see momentarily that those terms vanish in the decoupling limit. 

In the abelian case where $N=1$, this Yukawa simplifies to the form
\bea
\bar\Th\g_\m\left[X^\m,\Th\right] &=& {1\over 2 \pi} \int d\s\left(iL\bar\T \g^zD_1\T+\frac{i\la}{2}\left(\bar\T\g^{+-x}\T D_1x^++\bar\T \g^{+xi}\T D_1x^i\right.\right.\non\\
&& \left.\vphantom{\frac{i\la}{2}}\left.+2x^+\bar\T \g^xD_1\T-2x\bar\T \g^+D_1\T\right)\right).
\eea
It is worth noting that only couplings involving two fermions appear in the action. No four fermion couplings are generated. This is natural since since we are considering a smooth quotient of flat space which has vanishing curvatures.

%%%%%%%%%%%%%%%%%%%%%%%%%%%%%%%%%%%%%%%%%% 

\subsubsection{The extension to quantum mechanics}

Let us recall the procedure which we followed in the boson sector 
to go from the instanton action to the Lorentzian theory describing D0-branes.  
We let our ten matrices $X^\m_{mn}$ be functions of $\tau$ 
and we added a kinetic term $D_0X_\m D_0X^\m$ to the action.  
This term was reasonable because it was invariant under the 
ten-dimensional Poincar\'e symmetry, if we assumed that 
$\tau$ itself did not transform under this symmetry.  
Finally we gauge-fixed one of the coordinates
\be
 X^+_{mn}={\tau\over \sqrt 2 } \dd_{mn}.
 \ee  
 This procedure has the advantage of being completely covariant 
until the final step, while also reproducing the correct couplings 
in all static cases. Note that $\tau$ like $\s$ and $L$ 
is currently dimensionless. We will restore all dimensionful factors below.
%in section~\ref{parameters}.

Let us try to do the same with the fermions. 
We first note that in the quantum mechanics describing 
the D0-branes moving on the quotient space prior 
to Fourier transforming, the fermion kinetic term descends from 
the ambient flat space expression. All of the complications arising 
from the Fourier transform are captured by the 
Yukawa coupling~\C{yukawa}. We can then apply the decoupling limit 
to~\C{yukawa}\ which amounts to scaling 
$\lambda \sim \e$ and dropping terms that vanish as $\e\rr 0$. 
Let us call the $1+1$-dimensional fermion $\f$. 
The proposed $1+1$-dimensional action after decoupling is given by
\bea \label{complete}
S &=& \frac{1}{g_s}\int d\tau d\s\Tr\left(\vphantom{\frac{i\tau}{\sqrt 2}}(D_0x^i)^2 +(D_0x)^2-L^2 (D_1x^i)^2 -L^2 (D_1x)^2  - \sqrt{2} \lambda (x - \tau D_0x) F \right.\non\\
&& \left.+ (L^2+\hlf \lambda^{2}\tau^2) F^2 + {1\over 2} \left[ x^i, x^j\right]^2 + ( \left[x,x^i\right] +{i\tau\lambda \over \sqrt{2}} D_1x^i)^2  +i \bar\f \g^{0}D_0\f+iL\bar\f \g^zD_1\f \right.\non\\
&& \left.+\bar\f \left[\g^x\hat x+\g^ix^i,\f \right]-\frac{i\tau \lambda }{\sqrt 2}D_1\bar\f \g^x\f \right)
\eea
where $i=1, \ldots, 7$ and we have set $\ell_{s}=1$.  In the abelian case, 
the action simplifies to become
\bea\label{abaction}
S &=& \frac{1}{g_s}\int d\tau d\s \, \left((\p_0x^i)^2+(\p_0x)^2-(L^2+\hlf \lambda^{2}\tau^2)(\p_1x^i)^2-L^2(\p_1x)^2-\sqrt{2} \lambda (x-\tau\p_0x)F\right.\non\\
&& \left.+(L^2+\hlf \lambda^{2} \tau^2)F^2+i\bar\f \g^{0}\p_0\f +i\bar\f (L\g^z+\frac{\tau\lambda}{\sqrt 2}\g^x)\p_1\f\right).
\eea
Note that the matrix $L\g^z+(\tau \lambda/\sqrt 2)\g^x$ 
squares to give $(L^2+\hlf \lambda^{2}\tau^2)$ times the identity matrix. 
It should also be possible to derive the fermion couplings from the 
DBI action taking carefully account of the $B$-field and dilaton 
along the lines described 
in~\cite{Marolf:2003vf, Marolf:2004jb, Martucci:2005rb}.

\section{The Leading Quantum Corrections}

We now turn to quantum corrections to the effective dynamics.
The moduli space for the 
$1+1$-dimensional theory defined by~\C{complete}\ 
is parametrized by a choice of vacuum expectation value 
for $(x^{i}, x)$ and a choice of Wilson line in the $\sigma$ direction. We will put aside the choice of Wilson line for the moment. 

 For example,  for gauge group $SU(2)$ we parametrize the moduli space 
by the $Spin(7)$ and gauge-invariant combinations 
\be\label{vacuum}
\langle \Tr (x^{i})^{2}\rangle = {b^{2} \over 2}, \qquad \langle \Tr (x^{2}) \rangle = {c^{2} \over 2}.
\ee    
The choice of a $U(2)$ gauge group describes the interactions 
of two gravitons moving on the null-brane space-time. 
Restricting to $SU(2)$ factors out the center of mass motion.  

Now a glance at~\C{complete}\ shows that the $g_{s}$ 
plays the role of $\hbar$. We will compute 
the static potential energy as a function of $(b,c)$ to leading order 
in $g_{s}$. To do so, we employ the background field method. 

%%%%%%%%%%%%%%%%%%%%%%%%%%%%%%%%%%%%%%%%%%

\subsection{Gauge-fixing the action}

The first step needed to compute the quantum corrections to the action 
is to fix the gauge symmetry, and determine the ghost content 
of the theory. We need to implement a background field gauge condition. 
The usual gauge-fixing condition used in~\cite{Becker:1997wh}\ 
for D0-branes in flat space is
\be
G=-i \dot A_0+\eta_{\m\n}\left[X^\m_{b},X^\n\right] =0
\ee
where $(A_{0}, X^{\n})$ are fluctuating fields while 
$X^\m_{b}$ are background fields.  This choice of gauge fixing term 
ensures that the resulting quantum effective action 
is invariant under gauge transformations of the background fields.

Into this expression,  we insert the operators appearing 
in~\C{bigoperators}\ after taking the decoupling limit. 
The resulting gauge-fixing term takes the form
\bea \label{fixed}
G &= & -i\dot A_0+ i\La A_{1}'+\frac{i\tau \lambda}{\sqrt 2}x'
+\left[x^i_b,x^i\right]+\left[x_b,x\right]+\frac{\tau\lambda}{\sqrt 2 }
\left[x_b,A_{1}\right] \non\\ && +\frac{\tau\lambda}{\sqrt 2 }
\left[(A_{1})_b,x\right]+ \left[(A_{0})_b, A_{0}\right]
+\La\left[(A_{1})_b, A_{1}\right],
\eea
where $\La = L^{2} + {1\over 2}\lambda^{2}\tau^{2}$. 
Here a dot represents differentiation with respect to $\tau$, 
while a prime indicates differentiation with respect to $\s$.

The ghost action is obtained from the variation of $G$ 
with respect to a gauge transformation acting 
on the fluctuating fields.   Explicitly for the case at hand, 
the infinitessimal transformations with gauge parameter $\omega$ are
\bea
\dd x^i &=& i\left[\o,x^i+x^i_b\right],\non\\
\dd x &=& i\left[\o,x+x_b\right],\\
\dd A_{1} &=& \o'+i\left[\o,A_{1}+(A_{1})_b\right],\non\\
\dd A_0 &=& \dot\o+i\left[\o,A_0 + (A_{0})_{b}\right].\non
\eea
{}Finally one also adds a gauge-fixing term to the action~\C{complete}\ given by
\be
S_{\mathrm{gauge-fixing}}= {1\over g_{s}} \int d\tau d\s\, \Tr\left(G^2\right).
\ee

%%%%%%%%%%%%%%%%%%%%%%%%%%%%%%%%%%%%%%%%%%

\subsection{Warm-up: the abelian case}

Our real intention is to compute the effective potential 
for the moduli fields $b$ and $c$.  However, as a warm-up exercise 
let us consider the abelian case where we expand around 
the constant background configuration
\be
x_b=x_0,\quad  x^i_b=x^i_0,\quad  (A_{0})_b=0, \quad (A_{1})_{b}=0.
\ee
In this case we use the gauge-fixing term,
\be
G=-i\dot A_0+ {i\La} A_{1}'+\frac{i\tau\lambda}{\sqrt 2}x',
\ee
which varies under a gauge transformation in the following way:
\be
\dd G=i\left(-\ddot\o+\La\o''\right). 
\ee
This leads to the ghost action
\be\label{abeliangh}
S_{\mathrm{ghost}}={1\over g_{s}} \int d\tau d\s \, 
C^\ast\left(-\p_0^2+\La\p_1^2\right)C.
\ee
This, in part, explains the time-dependence appearing 
in the gauge-fixing term~\C{fixed}. The ghosts, like the 
gauge fields, see a time-dependent $\s$ circle. 

To obtain the other pieces of the action, 
we replace $x^\m$ by $x^\m+x_0^\m$ in~\C{complete}\ 
and add the gauge-fixing term.  The result is
\bea \label{gfabelian}
S &=& {1\over g_{s}} \int d\tau d\s
\left\{\left(\dot x^i\right)^2+\dot x^2-\dot A_0^2+2\La 
\dot A_0 A_{1}'+\sqrt 2 \tau\lambda \dot A_0 x'
-\La\left(x^{i\, \prime}\right)^2-\La\left(x'\right)^2
-\La^2 \left( A_{1} '\right)^2\right.\non\\
&& \left.- \sqrt 2 \tau \lambda \La A_{1}' x'
- \sqrt{2}  \lambda \left(x+x_0-\tau\dot x\right)F+\La F^2
+i\bar\psi\g^0\dot\psi+i\bar\psi\left(L\g^z
+\frac{\tau\lambda}{\sqrt 2}\g^x\right)\psi'\right\}
\eea
where
\be
F=\dot{A}_{1}-A_0'.
\ee
In this abelian case, the coupling dependence appearing 
in~\C{gfabelian}\ is irrelevant so we will drop the $1/g_{s}$ factor.

The bosonic part of this action has non-standard kinetic terms.  
To correct this,  we must perform a field redefinition.  Define 
\be
\tilde x = x+\frac{\tau \lambda}{\sqrt 2} A_{1}. 
\ee
This definition does not respect gauge-invariance but since 
we have already gauge-fixed, this is not a problem. 
With these redefinitions, the bosonic action simplifies to
\bea \label{adjact}
S= \int d\tau d\s\left\{  \left(\dot x^i\right)^2
+\dot{\tilde x}^2+\dot{A}_{1}^2 - \dot A_0^2 
-\La\left[\left(x^{i\prime}\right)^2+\left(\tilde x'\right)^2
+\left(A_{1}'\right)^2 -\left(A_0'\right)^2\right]  
+ 2\sqrt 2 \lambda  \tilde x A_0'\right\} . \,
\eea

We decompose the fields into their Fourier modes labeled by $n$. 
At each level, we find eight bosons of ``mass" $m^2=-n^2(L^2+\lambda^{2}\tau^2/2)$ 
with conventional kinetic terms. The operator governing quadratic
fluctuations is
\be
H = -\p_{\tau}^{2} - n^2 \left(L^2+{ \tau^2 \lambda^{2} \over 2} \right)\ .
\ee
Integrating over 
the quadratic fluctuations for each such boson generates a factor, 
\be
\exp \left(  i \int d\tau d\s  \, V_{\rm eff} \right) 
= {\det}^{-1/2}(H) = {\det}^{-1/2} \left( - \p_{\tau}^{2} 
- n^2(L^2+ { \tau^2 \lambda^{2} \over 2}) \right). 
\ee
To express this as an effective potential, we use a proper time 
representation for the determinant,
\be\label{defdet}
 {\det}^{-1/2}(H)  = \exp \left( {1\over 2} 
\int {d\tau d\s \over 2\pi}  \int \frac{dt}{t} {e^{- it (H - i\e)}} (\tau, \tau)\right),
\ee 
where $t$ is the Lorentzian proper time, and an $i\e$ is inserted for convergence. The integral over $\s$ 
is not part of the trace but can be inserted harmlessly in this 
case because $\s$ does not appear in the kernel. 

The only 
information we require is the propagator for $H$, 
given by a continuation of Mehler's formula for the simple harmonic oscillator
\bea\label{Mehler}
e^{-itH}(\tau_1, \tau_2) &=& e^{in^2 L^2 t} \left( 
{ in \lambda \over 2 \pi \sqrt{2} 
\sinh(\sqrt{2} n\lambda t)}\right)^{1/2} 
\\
& &\hskip 2cm
\times\exp\left[  {in\lambda \over \sqrt{2} 
\sinh(\sqrt{2} n\lambda t)} \left(\coeff12(\tau_1^2+\tau_2^2) \cosh(\sqrt{2} n\lambda t) 
- \tau_1\tau_2 \right) \right]\ .
\non
\eea
The large $t$ behavior of~\C{defdet}\ corresponds to 
the infrared contribution to the potential. From~\C{Mehler}, we 
see that this contribution is highly suppressed by both 
the $n^{2} L^{2}$ mass term, the time-dependent 
$n\tau^{2}$ term, and the $\sinh(\sqrt{2} n\lambda t)$ prefactor. This is what we expect for a massive particle with a growing time-dependent mass.  

%More interesting is the small $t$ ultraviolet behavior of the proper 
%time integral~\C{defdet}. 
%For small $t$, the exponential factors 
%appearing in~\C{Mehler}\ can be approximated by the sum 
%\be\label{uvdiv}
%\sum_{n=0}^{\infty}  e^{- n^{2} t \La} \sim (t \La)^{-1/2}. 
%\ee
%So the effective potential diverges for small $t$ in the following way:
%\be\label{potdiv}
%V_{\rm eff} \sim \int dt \, \lim_{t\rr 0} \, {e^{-t H} \over t} 
%(\tau, \tau) \rr \int dt \left( {1\over t^{2} \sqrt{\La}} + \ldots\right) .
%\ee
%Up to the time-dependent $\La$ factor, this is precisely 
%the ultraviolet divergence we expect in a two-dimensional field theory. 

Now there are two additional bosons $\tilde x$ and $A_{0}$. This is a coupled system that is more subtle to analyze. Path-integrating over $\tilde x$ and $A_{0}$ results in the determinant, 
\be\label{harderdet}
{\det}^{-1/2}( H^{2} + q^{2})
\ee
where $q^{2} = 2n^{2} \lambda^{2}$.  Note that the spectrum of $H^{2} + q^{2}$ is well-behaved because $H$ is Hermitian, and gapped because of $q$.  
There is a convenient way to represent the determinant~\C{harderdet}\ using the product, 
\be\label{factor}
{\det}^{-1/2}( H + iq) \times {\det}^{-1/2}( H - iq). 
\ee
The reason this is convenient is that we can directly use the kernel~\C{Mehler}\ to represent the effective potential. This results in complex masses, 
\be m^2=n^2(L^2+{\tau^2 \lambda^{2}\over 2})\pm\sqrt 2 in \lambda, \ee
which really require different proper time representations~\C{defdet}\ (differing in the sign of $t$) depending on the sign 
of the imaginary part of the mass. We will avoid this issue and directly compute the determinant~\C{harderdet}.

The complex ghosts with action~\C{abeliangh}\ simply cancel the 
contribution of $2$  of the $8$ bosons with mass 
$m^2=n^2(L^2+\tau^2 \lambda^{2}/2)$. We are left with the fermions. 
To find the fermion mass spectrum, it is convenient to introduce complex masses much as we could have done  in the
$(\tilde{x}, A_{0})$ system. The path-integral over the real fermions gives the Pfaffian of the operator, 
\be\label{pfaf}
-i \p_{\tau} + i \g^{0} \left( L \g^{z} +\frac{\tau\lambda}{\sqrt 2}\g^x \right) \p_{\sigma}.
\ee 
This is a Hermitian operator with real eigenvalues. It is more convenient to rewrite this Pfaffian as $ \sqrt{ \det (\Dslash) }$ in terms of the Dirac operator. The square of the Dirac operator unlike the square of~\C{pfaf}\ takes a nice form: 
\be
 \left(i \g^0\p_\tau+ i\left(L\g^z + \frac{\tau \lambda}{\sqrt 2}\g^x\right)
\p_\s\right)^2=\p_\tau^2- \La\p_\s^2 - {\lambda\over \sqrt 2}\g^0\g^x\p_\s.
\ee
The eigenvalues of the matrix $n^2\La - (in \lambda/\sqrt 2)\g^0\g^x$ are complex:  
$n^2\La+in\lambda /\sqrt 2$ and 
$n^2\La-in\lambda /\sqrt 2$. The $16$ real fermions therefore 
contribute for each $n$ a factor
\be\label{fermdet}
{\det}^{4} \left[ \p_\tau^2+n^2\La-{in \lambda / \sqrt 2} \right] 
\times {\det}^{4} \left[ \p_\tau^2+n^2\La+ {in \lambda / \sqrt 2} \right]
\ee
which we express as
\be
{\det}^{4} \left[ H^{2} + q^{2}/4 \right]. 
\ee

Collecting together these results, we need to compute
\be\label{whatweneed}
{\det}^{-1/2} \left[ { H^{6}(H^{2} + q^{2}) \over \left(H^{2}+ q^{2}/4 \right)^{4} }\right] = \exp \Tr \left[ 2 \log (H^{2}+ q^{2}/4) -{1\over 2} \log (H^{6}(H^{2} + q^{2})) \right]
\ee
to determine the potential. We will expand the logarithms formally in inverse powers of $H$, 
\be\label{expansion}
2 \log (H^{2}+ q^{2}/4) -{1\over 2} \log (H^{6}(H^{2} + q^{2})) = {3\over 16} {q^{4} \over H^{4}} - {5\over 32}{q^{6} \over H^{6}} +{63\over 512} {q^{8} \over H^{8}} - {51\over 512} {q^{10} \over H^{10}} + \ldots.
\ee
Now we can represent inverse powers of $H$ using a Lorentzian proper time formalism
\be
\left({1\over H}\right)^{n} (\tau,\tau) = {(i)^{n} \over (n-1)!} \int dt \, t^{n-1} e^{-it (H-i\e)} (\tau, \tau).
\ee
 This amount to the replacement
 \be
 \left({1\over H}\right)^{n} \, \rr \, {1\over t} { (it)^{n} \over (n-1)!}
 \ee
in the series~\C{expansion}. On substituting, we find the series
\be
{1\over 32 t} \left(  {(q t)^{4}} + {1\over 24}{ (q t)^{6}} +{1\over 1280} {(q t)^{8}} +  {17\over 1935360} {(qt)^{10} } + \ldots \right) = {8\over t} \sinh^{4} \left({q t \over 4} \right).
\ee
This results in the following Lorentzian proper time expression for the effective potential, 
\bea\label{Lorv}
iV_{\rm eff} 
&=& {1\over 2\pi} \int \frac{dt}{t} \sum_{n=0}^\infty
8 \sinh^{4} \left({n \lambda t \over 2 \sqrt{2}} \right) e^{i n^2L^2 t} 
\non\\
& &
\hskip 2cm
\times\Bigl( \frac{i n\lambda}{2\pi\sqrt2\sinh \left(\sqrt2 n\lambda t \right)}\Bigr)^{1/2}
\,\exp\Bigl[ \frac{i n\lambda\tau^2}{\sqrt2}\,
\tanh\left(\coeff{n\lambda t}{\sqrt2}\right)\Bigr]. 
\eea
Note that this potential~\C{Lorv}\ has an apparent large $t$ infrared divergence which we will revisit after rotating to Euclidean space. 

There are a few points worth noting. Let us rewrite the $ 8 \sinh^{4} \left({n \lambda t \over 2 \sqrt{2}} \right)$ term in the form
\be
{1\over 2} \Bigl[ 8-2+2\cosh (\sqrt2 n\lambda t)-8\cosh \left(\coeff{n\lambda t}{\sqrt2}\right)\Bigr].
\ee
This is the collection of prefactors we would have obtained had we directly represented~\C{factor}\ and~\C{fermdet}\ using the same proper time representation regardless of the sign of the imaginary part of the mass.  If we restrict to 
$n=0$, we see that there are effectively $8+2-2$ net 
bosonic determinants which are precisely canceled by 
the fermionic determinants. 
This reflects the underlying 
supersymmetry present in the $n=0$ sector. 

Similarly, the leading UV divergence also cancels in 
the effective potential. This comes about because the $\cosh$ prefactors that distinguish the fermion and 
$(\tilde x, A_{0})$ effective potential terms do 
not change the leading $t\rr 0$ behavior of
the determinants; the 
underlying supersymmetry still kills the $1/t^{2}$ 
divergence characteristic of $1+1$-dimensional field theory in the proper time integral in~\C{defdet}. 

We can now rotate~\C{Lorv}\ to both Euclidean proper time and Euclidean time sending
\be
t \rr  it, \qquad \lambda \rr -i\lambda, \qquad \tau \rr i \tau.
\ee
The reason we need to rotate $\lambda$ is to ensure that the $\sinh$ factors in~\C{Lorv}\ continue in a sensible way to Euclidean space. One finds
\bea
iV_{\rm eff} 
&=& {4\over \pi} \int\frac{ds}{s} \sum_{n=0}^\infty  \sinh^4\left(\frac{s}{2\sqrt2}\right)\,
e^{-nL^2 s/\lambda}\left( \frac{n\lambda}{2\pi\sqrt2\sinh \left(\sqrt2 s \right)}\right)^{1/2} \non \\
& &
\hskip 2cm
\times \exp\Bigl[-\frac{n\lambda\tau^2}{\sqrt2}\,\tanh\left(\frac{s}{\sqrt2}\right)\Bigr], 
\label{effpotl} 
\eea
where we have defined $s=n\lambda t$.

We see that either $n\tau^2\lambda\gg 1$ or
$nL^2/\lambda\gg 1$ forces $s\ll 1$; in other words,
$n\Lambda/\lambda\gg 1$ implies only $s\ll 1$
contributes to the integral.  In this regime, we have
\bea
\label{veffabelian}
2\pi V_{\rm eff} &\sim&  \int\frac{ds}{s} \; s^4\sum_{n=1}^\infty
e^{-n\Lambda s/\lambda} \Bigl(\frac{n\lambda}{\pi s}\Bigr)^{1/2}
\non\\
&\sim&  {1\over  \sqrt{\pi}} 
\zeta(3)\Gamma(7/2)\frac{\lambda^4}{\Lambda^{7/2}}
\eea
This result is easily understood from the D0-brane picture
as arising from the one-loop $v^4/r^7$ interaction of 
matrix theory, summed over images under the orbifold group.

When $s$ becomes sufficiently large, the effective potential~\C{effpotl}\ can have an infrared divergence. This occurs roughly when
\be
\label{couplescales}
n L^{2} \lesssim  \lambda / \sqrt{2}.
\ee
From~\C{adjact}, we see that the right hand side is the coupling constant in the abelian theory 
mixing the modes of $(\tilde x, A_{0})$,
so the divergence occurs when the energy of the lightest Kaluza-Klein mode 
is of order the scale set by the coupling constant. 
Strong coupling physics appears in the low energy regime of the matrix model,
where the dimensionful Yang-Mills coupling becomes effectively large.
The infrared divergence 
appears to herald the onset of strong coupling, 
and the breakdown of perturbation theory. 

Given that the ultraviolet contribution to the potential collapses to the $v^4/r^7$ interaction, it seems quite possible that the non-renormalization theorems for this and higher velocity interactions~\cite{Paban:1998ea}\ can be extended to this potential. 

The result~\pref{veffabelian} yields a time-dependent contribution 
to the energy of the center-of-mass dynamics of the system 
that depends on the orbifold parameters $(L,\lambda)$
but is independent of the center of mass position and velocity.
This energy indicates the time-dependent scale of supersymmetry breaking 
in the system.

%Indeed the proper time integral is finite as $t \rr 0$. 
%To see this we sum over all the contributions which gives an 
%effective integrand in terms of Mehler's kernel~\C{Mehler}
%\bea
%V_{\rm eff} & = & \hlf  \int dt \, \sum_n=0^\infty
%%\lim_{t\rr 0} \, 
%\sum_{n=0}^{\infty} \left( 8 -2 + 2\cos (\sqrt{2} nt) - 
%8 \cos({nt\over \sqrt{2}}) \right) {e^{-t H} \over t} (\tau, \tau) 
%\non\\ 
%&\sim & \hlf  \int dt \, \sum_{n=0}^{\infty} 
%\left( {n^{4} t^{4} \over 4} + \ldots \right) 
%\left( {1\over t^{3/2}}   e^{- n^{2} t \La}  + \ldots \right) 
%\non\\
%&\sim & \hlf \int dt \, \left({ t^{4} \over 4} 
%+ \ldots \right)  \left( {1\over t^{4} \La^{5/2}} + \ldots\right). 
%\eea   
%Note that the potential has an overall sign as if 
%it were effectively generated by bosons. 
%Now in this abelian case, none of these determinants depend on the 
%moduli and so we usually discard them. Fortunately, the computations we 
%require in the non-abelian case are very similar in spirit.     

%%%%%%%%%%%%%%%%%%%%%%%%%%%%%%%%%%%%%%%%%%

\subsection{Particle production}

A time-dependent background generically results in particle production.
While there is no particle production in the null-brane  vacuum 
because of the existence of a null killing vector, 
there is particle production in the matrix description. 
This is because the matrix theory describes objects 
carrying longitudinal momentum which preserves a different set of supersymmetries than the null-brane background.

The linearized dynamics of the bosonic fields is 
solved by parabolic cylinder functions
(our discussion here follows~\cite{Brout:1990ci}),
\bea
	& &[-\partial_\tau^2-\omega^2\tau^2]\psi_\nu(\tau) = \nu\psi_\nu(\tau),
\label{parcyl}\\
	& &\psi_\nu = \alpha\,\frac{e^{-\mu\pi/4}}{(2\omega)^{1/4}}
		\;D_{-i\mu-\frac12}\Bigl( e^{\frac{i\pi}{4}}\sqrt{2\omega}\,\tau\Bigr)
		\ +\ \beta\,\frac{e^{-\mu\pi/4}}{(2\omega)^{1/4}}
		\;D^*_{-i\mu-\frac12}\Bigl( e^{\frac{i\pi}{4}}\sqrt{2\omega}\,\tau\Bigr),
\nonumber
\eea
where the Bogolubov coefficients $(\alpha, \beta)$ satisfy $|\alpha|^2-|\beta|^2=1$, and
\be
	\mu = \frac{\nu}{2\omega} = \frac{nL^2}{\sqrt 2 \la}.
\ee
We use the values $\omega = n\lambda/\sqrt 2$ and $\nu= n^2L^2$ relevant
to the dynamics of the $n^{\rm th}$ mode of $x^i$.
Note that $\mu$ is simply the dimensionless ratio of the energy cost
and the coupling of a given mode, see equation~\pref{couplescales}.

The parabolic cylinder functions have the asymptotics
\be
	D_{-i\mu-\half}\Bigl( e^{\frac{i\pi}{4}}\sqrt{2\omega}\,\tau\Bigr) \sim
\begin{cases}
\frac{e^{\mu\pi/4}}{\sqrt{2\omega\tau}}
	e^{-i(\frac\omega2\tau^2+\mu\log(\sqrt\omega\tau))}
	& \tau\to+\infty \cr
	& \cr
\frac{e^{\mu\pi/4}}{\sqrt{2\omega|\tau|}}
\Bigl[ie^{-\pi\mu}e^{-i(\frac\omega2\tau^2+\mu\log(\sqrt\omega|\tau|))}
	& \cr
\qquad\qquad +e^{i\gamma}(1+e^{-2\pi\mu})^\half
e^{+i(\frac\omega2\tau^2+\mu\log(\sqrt\omega|\tau|))}\Bigr]
		& \tau\to-\infty \cr
\end{cases}
\label{parcylasymp}
\ee
where $\gamma$ is the phase
\be
\gamma =-\frac{i}{2} \log \Biggl[
{\Gamma(\half-i\mu)\over \Gamma(\half+i\mu)}\Biggr]+\mu\log 2.
\label{phasechange}
\ee
%{\bf overall sign here??? check it...}
The Bogolubov coefficient $\beta$ expressing the probability amplitude for 
particle production may be read from~\pref{parcylasymp}:
\be
|\beta| = e^{-\pi\mu}=\exp\Bigl[-\frac{\pi nL^2}{\sqrt 2\lambda}\Bigr].
\label{bogolcoeff}
\ee
Thus the mode production rate becomes of order one as the effective coupling
of a mode becomes of order one.  This feature is related to the 
onset of infrared divergences in the one-loop determinants calculated
in the previous subsection; the infrared singularities and the copious mode production
are both signals of the breakdown of the weak-coupling perturbative expansion.

The above calculation of mode production is nicely reproduced 
from the underlying picture of D0-branes on the null-brane orbifold.  
Consider a D0-brane and its image under the orbifold group~\pref{quotgroup}.
They undergo a scattering process which at 
leading order in string perturbation theory 
was calculated in~\cite{Bachas:1995kx}.
The real part of this annulus amplitude 
can be interpreted as the eikonal phase shift of
scattering D-branes
(see for instance~\cite{Bachas:1995kx,Douglas:1996yp}); 
the imaginary part of the phase shift gives the
probability of vacuum decay.  
Alternatively, the imaginary part
is the pair production probability of open strings 
stretching between the branes via the optical theorem.  
These stretched strings are the Fourier modes in the T-dual 
field theory, and so the production amplitude
of these stretched strings should agree 
with the field theoretic calculation above.

In the matrix theory limit,
the annulus amplitude becomes the one-loop amplitude
in the field theory~\pref{complete}, 
leading to the collection of determinants
evaluated in the previous subsection
(integrated over $\tau$).
The result of~\cite{Bachas:1995kx,Douglas:1996yp} 
for the scattering phase shift of
a D0-brane and its image under $g^n$ is
\be
\label{phaseshift}
\d_n=-\hlf\int_0^\infty \frac{ds}{s}e^{-snL^2/\lambda}
\frac{1}{\sin( s/\sqrt 2)}\left(4\cos( s/\sqrt 2)-\cos(\sqrt 2  s)-3\right)
\ee
\be
e^{i\d_n}=\tanh^2\left(
\frac{\pi nL^2}{\sqrt 2 \la}\right)
\left[\frac{\sqrt 2 inL^2+\la}{\sqrt 2 
inL^2-\la}\right]^\hlf \left[\frac{\G\left(
-\frac{inL^2}{\sqrt 2 \la}\right)\G\left(
\hlf+\frac{inL^2}{\sqrt 2 \la}\right)}{\G\left(\frac{inL^2}{\sqrt 2 \la}
\right)\G\left(\hlf-\frac{inL^2}{\sqrt 2 \la}\right)}\right]^2.
\label{strcreate}
\ee
The result~\pref{phaseshift} is simply the
local  expression for the determinants in $\tau$, given in equation~\pref{effpotl},
integrated over $\tau$ and continued back in $\lambda$.
The full phase shift sums this result for $\delta_n$ over the images
labelled by $n$, which is the same as the sum over modes
in the T-dual field theory.

%In particular, we have the contribution of the $n^{\rm th}$ mode
%\be
%\Im[\delta_n] = \sum_{k=1}^\infty\frac1k e^{-b^2k/v} \cdot 8[1-(-)^k]
%  	= 8\log\Bigl[\frac{1+e^{-b^2/v}}{1-e^{-b^2/v}}\Bigr]
%\label{strcreate}
%\ee
%where $b$, $v$ is the impact parameter and relative velocity, respectively.
%The factor of eight is the number of polarization states, for which the
%bosons contribute the numerator in the log 
%and the fermions give the denominator.
%%$\log[1+e^{-b^2/2v}]$ and the fermions $-\log[1-e^{-b^2/2v}]$.
%{\bf need to square notation and numerical factors???}

How is this decay probability related 
to the Bogolubov coefficient calculated above?
%It was noted in~\cite{Bachas:1995kx}
%that the result \pref{strcreate} 
%for the imaginary part of $\delta$ 
%is identical to the probability of decay of the vacuum 
%by charged particle pair production in a constant
%background electric field; the wave equation for a charged particle in
%a constant electric field is precisely \pref{parcyl}.
%In more detail,
The decay probability is the overlap of the in- and out-vacua, 
$|_{\rm \,out}\!\langle 0|0\rangle_{\rm in}|^2$.  The Bogolubov transformation 
of (uncharged) bosonic creation/annihilation operators
\be
  a_{\rm out} = \alpha a_{\rm in} + \beta a^\dagger_{\rm in}
\ee
(with $|\alpha|^2-|\beta|^2=1$) 
allows one to write the out vacuum as
\be
\ket{0}_{\rm out} = \frac{1}{\sqrt{\alpha}}\, \exp\Bigl[-\frac{\beta}{2\alpha}
	(a_{\rm in}^\dagger)^2\Bigr] \ket{0}_{\rm in}
\ee
and thus 
\be
-\log \Bigl(\bigl |\!\,_{\rm \,out}\!\langle 0|0\rangle_{\rm in}
\bigr |^2\Bigr)=\log |\alpha| = \coeff12\log[1+|\beta|^2].
\ee
The analogous calculation for fermions yields, \be -\coeff12\log[1-|\beta|^2], \ee
because in this case the Bogolubov coefficients 
are related by $|\alpha|^2+|\beta|^2=1$.  
Finally,  \be \log \left(\frac{1+|\beta|^2}{1-|\beta|^2} \right)=-\log\tanh(\log|\beta|^{-1}) \ee
connects the decay probability to~\pref{strcreate}.
Note also that we can understand
the real part of the phase shift $\delta$ as
the phase $\gamma$, appearing in~\pref{phasechange},
accumulated in propagating over the inverted oscillator barrier
(taking into account the frequency shifts involved for
the various bosons and fermions).

This relation between the two results \pref{bogolcoeff} and \pref{strcreate} 
yields $\beta=e^{-\pi\mu}=e^{-\pi b^2/2v}$.
The group element $g^n$ of the orbifold generates an image D0-brane
whose impact parameter is $b=nL$ and relative velocity $v=n\lambda/\sqrt 2$.

%
%{\bf need to check coefficients!!!???  still off by some 2's and $\pi$'s} 

We see in the result~\pref{bogolcoeff} the effect of the order of limits,
and the various scales.  If we send $L\to 0^+$ at fixed $\lambda$,
mode production is of order one for all the $N^2$ modes
of the system.  The dynamics after passing through the neck of the geometry
involves energies that grow with $N$ and will not have a uniform large $N$ 
limit.  On the other hand, this is a DLCQ artifact.  Recall from our discussion in section~\ref{largeN}\ that $\lambda\sim 1/N$
because it is a velocity in the boosted frame of the matrix model.

Taking the large $N$ limit first, particle production is an effect
suppressed by $e^{-cN}$ for some constant $c$, and thus totally
unimportant.  Note, however, that strings that thread through $k\sim N$
D0-branes have an action cost that is not suppressed parametrically
in $N$; these strings will be produced in the passage through the neck,
however this effect is a collective multi-particle excitation from the
point of view of the perturbative matrix model, not governed
directly by the production rate~\pref{bogolcoeff}.  It is an open question
whether such collective excitations are well-behaved in 
the large $N$ limit.  The answer to this question is central to
the issue of whether matrix theory gives a well-defined answer
to what happens at the cosmological singularity at small $L$ and large $N$.

%%%%%%%%%%%%%%%%%%%%%%%%%%%%%%%%%%%%%%%%%%

\subsection{The non-abelian theory}
\label{nonabelian}

Now let us to the case of a broken $U(2)$ gauge group. We expand around the vacuum configuration~\C{vacuum}\ generalized to permit velocities with the explicit choice of background fields
\be
  x_b^i=\frac{b^i+\beta^i\tau}{2}\s^3, \quad x_b=\frac{c+\gamma\tau-\frac{\la d}{\sqrt 2L}\tau^2}{2}\s^3,\quad (A_{1})_b= \frac{d+\delta\tau}{2} \s_3, \quad  (A_{0})_{b}=0,
\ee
where $\s^a$ are the Pauli matrices. 
%The parameter $d$ is periodic and corresponds in the D0-brane picture to the position of the brane in the $z$ direction. The corresponding momentum is the electric flux, $\dd$, which is quantized as usual.  
The unusual form of $x_b$ is required to satisfy the classical equations of motion which include mixing of $x$ and $F$.  However, if we again make a change of variables to $\tilde x=x+\la\tau A_1/\sqrt 2$, then we have the simpler setup
\be
{\tilde x}_b=\frac{c+\tilde\gamma\tau}{2}\s^3,\quad \tilde\gamma=\gamma+\frac{\la d}{\sqrt 2}.
\ee
Since the dependence on the relative velocities $\gamma$, $\beta^i$,
and $\delta$
follows from Galilean invariance, we suppress the velocity
dependence for the moment but will restore it later.
We then consider the fluctuations around this background, 
\be
x^i_n= \hlf\left(x^{i\,0}_n{\bf{1}}+ x^{i\,a}_n\s^a\right),\ {\mathrm{etc.}}
\ee
As usual, we will we ignore the $U(1)$ center of mass physics and focus on the interacting non-abelian theory. 

We will repeat the procedure used in the abelian case except we will work directly with complex masses. We find a collection of particles with masses conveniently parametrized in terms of
\be\label{defr}
r_\pm^2=L^2(d\pm n)^2+{b}^2+\left(c\pm\frac{n \lambda \tau}{\sqrt 2}\right)^2.
\ee
{}For each $n$, there are $8$ massive bosons with masses, $m^{2} = r_{+}^{2}$ and $8$ with masses  $m^{2} = r_{-}^{2}$. There is also one complex ghost with  $m^{2} = r_{+}^{2}$ and one with  $m^{2} = r_{-}^{2}$. The shift in $\tau$ induced by $c$ in~\C{defr}\ is physically significant when we allow the moduli fields to vary slowly as functions of space-time. 

In addition to these particles, there are again a pair of real bosons with $m^{2} = r_{+}^{2} \pm \sqrt{2} in \lambda$ and a pair of bosons with $m^{2} = r_{-}^{2} \pm \sqrt{2} in \lambda$. The complex masses appear for exactly the same reasons as in the abelian case.  

Finally, the fermions also split into two groups with the first $16$ real fermions generating a factor of
\be\label{det1}
{\det}^{4} \left[ \p_\tau^2+r_{+}^{2}-{in \lambda / \sqrt 2} \right] \times {\det}^{4} \left[ \p_\tau^2+r_{+}^{2}+ {in\lambda / \sqrt 2} \right],
\ee
while the second $16$ real fermions generate the determinants
\be\label{det2}
{\det}^{4} \left[ \p_\tau^2+r_{-}^{2}-{in\lambda / \sqrt 2} \right] \times {\det}^{4} \left[ \p_\tau^2+r_{-}^{2}+ {in\lambda / \sqrt 2} \right].
\ee

The potential therefore splits into the sum of two contributions,
\be
V_{\rm eff} = V_{\rm eff}^{+} + V_{\rm eff}^{-}, 
\ee
where
\bea \label{finalpot}
V_{\rm eff}^{\pm} &=& {1\over 4 \pi} \int \frac{dt}{t} \, \sum_{n=0}^{\infty} \left(8 -2 + 2\cosh (\sqrt{2} n\lambda t) - 8 \cosh\left({n \lambda t\over \sqrt{2}} \right)\right) {K_\pm(t, { \sqrt{2} c \over n} \pm \lambda \tau, n)}, \non\\
&=& {4\over \pi} \int \frac{dt}{t} \, \sum_{n=0}^{\infty} \sinh^{4}\left({n \lambda t\over 2\sqrt{2}} \right) {K_\pm(t, { \sqrt{2} c \over n} \pm \lambda \tau, n)}, 
\eea
using Euclidean proper time, and 
\bea
K_\pm(t,\tau,n) =  e^{- ((n\pm d)^2 L^2+b^{2}) t} \left( {n \lambda \over 2 \pi \sqrt{2} \sinh(\sqrt{2} n \lambda t)}\right)^{1/2} \exp\left[ - {n\lambda \tau^2 \over \sqrt{2}} \tanh\left( {n\lambda t\over \sqrt{2}} \right) \right].
\eea
Note that from the discussion in the abelian case, 
we see that the potential is UV finite (\ie\ as $t\to 0$). There is an infrared 
divergence when 
\be
(n\pm d)^2 L^{2} + b^{2} \lesssim n \lambda/ \sqrt{2} \ .
\ee
As in the abelian case, this appears to signal the breakdown of perturbation theory.  
   
Once again, for $nr^2/v\gg 1$ the dominant contribution is from the
UV region $t\to 0$.  As in the abelian case, we can understand the
effective potential as due to the sum over orbifold group images
of $v_\pm^4/r_\pm^7$, where
\bea
v_\pm^2 &=&\vec \beta^2+(\tilde\gamma\pm \frac{n\lambda}{\sqrt 2})^2+\delta^2
\non\\
r_\pm^2 &=& (\vec b+\vec\beta\tau)^2+(c+\tilde\gamma\tau\pm\frac{n\lambda}{\sqrt 2}\tau)^2 + (d+\delta\tau\pm n)^2L^2.
\eea
and we have restored the full dependence on the velocities. We have been a little cavalier about the sign of the potential. However, from the D0-brane picture, we can see that a potential that originates from the $v^{4}/r^{7}$ interaction will have the same sign as the kinetic terms in the action~\cite{Liu:1997gk}; the potential is therefore attractive. 

\begin{figure}[h] 
\begin{center}
\[
\mbox{\begin{picture}(100,190)(100,110)
\includegraphics[scale=.5 ]{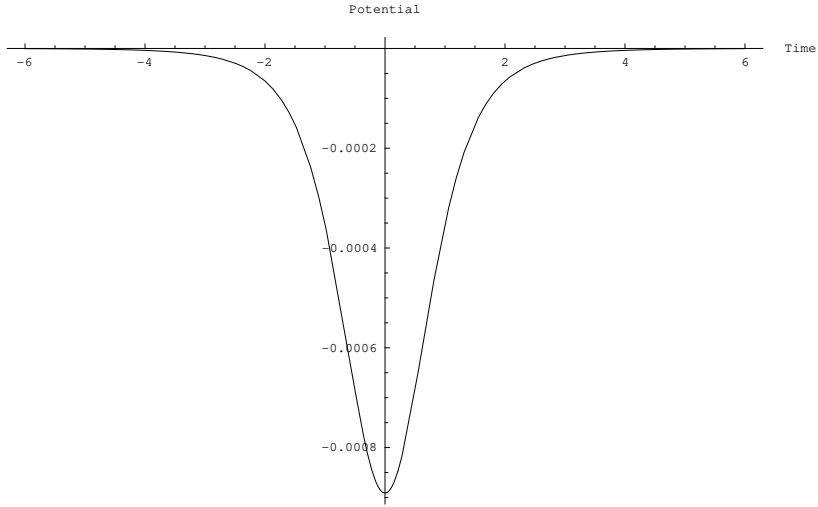}
\end{picture}}
\]
\caption{\small The potential as a function of time with $b=2$ and  $L=1$.}
\label{finfig}
\end{center}
\end{figure}

The potential is decaying rapidly with $|\tau|$. 
This implies that the flat directions are restored 
as $|\tau| \rr \infty$ and we recover the conventional picture 
of gravity in matrix theory, at least in terms of the structure 
of the potential. We can get some feel for the potential by setting $c=d=0$ along with the velocities $\beta^{i}=  \tilde{\gamma}=\delta=0$ and $\lambda=1$. The corresponding static potential as a function of $\tau$ appears in figure~\ref{finfig}. The static potential as a function of impact parameter for various choices of $L$ appears in figure~\ref{mygr2}.

\subsection{Final comments}

There are a few obvious generalizations of the above analysis. As we described briefly in section~\ref{parameters}, one can compactify additional dimensions. 
For instance, a 
toroidal compactification in $p$ additional directions
will extend the 1+1-dimensional field theory to a theory on $T^{p+1}$,
one of whose cycles has time-dependent proper size $\Lambda$. The most obviously interesting cases are $p=1$ and $p=2$ where we expect novel field theory phenomena to occur: for $p=1$, time-dependent instanton effects while for $p=2$ we expect to find a version of S-duality. 

However, there are other ways to compactify. We could have all the cycles undergo a bounce like the bounce in the $z$-direction by 
combining the translational identification with a boost 
as in~\pref{gen}; then each cycle has a time-dependent size
$(L_i^2+\frac12\lambda_i^2\tau^2)^{1/2}$.

An object passing through the neck of such a bouncing $T^p$ 
generates a singularity as we send the $L_i$ to zero. 
Again, for small but nonzero $L_i$, what initially forms is a black
$p$-brane filling the torus.  As the torus expands after the bounce,
this $p$-brane will experience a Gregory-Laflamme instability
and break up into a collection of black holes.
This result is reminiscent of the proposal of~\cite{Banks:2001px}
for a cosmology whose initial state is a dense gas of black holes.
Of course, in the present case the gas of black holes is anisotropic,
filling only those directions that are compactified;
we do not currently have a non-perturbative description of the situation when
all spatial directions are compact.

%%%%%%%%%%%%%%%%%%%%%%%%%%%%%%%%%%%%%%%%%%
%%%%%%%%%%%%%%%%%%%%%%%%%%%%%%%%%%%%%%%%%%

\
\section*{Acknowledgements}
 
The work of E.~M. is supported in part by DOE grant DE-FG02-90ER-40560. 
The work of D.~R. is supported in part by a Sidney Bloomenthal Fellowship.
The work of S.~S. is supported in part by NSF CAREER Grant No. PHY-0094328 
and by NSF Grant No. PHY-0401814.

%%%%%%%%%%%%%%%%%%%%%%%%%%%%%%%%%%%%%%%%%%
%%%%%%%%%%%%%%%%%%%%%%%%%%%%%%%%%%%%%%%%%%

%%%%%%%%%%%%%%%%%%%%%%%%%%%%%%%%%%%%%%%%%%
%%%%%%%%%%%%%%%%%%%%%%%%%%%%%%%%%%%%%%%%%%

\newpage
%%%%%%%%%%%%%%%%%%%%%%%%%%%%%%%%%%%%%%%%%%%%%%%%%%%%%%%%%%%%

%\bibliographystyle{utphys}
%\bibliography{myrefs}
\providecommand{\href}[2]{#2}\begingroup\raggedright\endgroup
\end{document}